%&cp-aa
\input epsf.tex
\voffset 1 true cm
\newdimen\tabledimen  \tabledimen=\hsize
\def\table#1#2{\tabledimen=\hsize \advance\tabledimen by -#1\relax
\divide\tabledimen by 2\relax\vskip 1pt
\moveright\tabledimen\vbox{\tabskip=1em plus 4em minus 0.9em
\halign to #1{#2}} }

\def\note #1]{{\bf #1]}}
\def\gwig{{\leavevmode\kern0.3em\raise.3ex\hbox{$>$}
\kern-0.8em\lower.7ex \hbox{$\sim$}\kern0.3em}}

\def\jcd{Christensen-Dalsgaard}
  % top rule
\def\tabmidrule{\noalign{\smallskip\hrule\smallskip}}             % middle rule
       % bottom rule
\def\eqname#1{\global\advance\eqnum by 1\relax
\xdef#1{{\noexpand{\rm}(\number\eqnum)}}
{\rm(\the\eqnum)}}
\def\viz{{\it viz.}}
\def\d{{\rm d}}
\def\Fsurf{F_{\rm surf}}

\def\eg{{\rm e.g.}}
\def\ie{{\rm i.e.}}
\def\cf{{\rm cf.}}
\def\etal{{\rm et al.}}

\def\jcd{Christensen-Dalsgaard}
\def\eg{{e.g.}}
\def\cf{{cf.}}
\def\ie{{i.e.}}
\def\etal{{et al.}}
\def\Fsurf{F_{\rm s}}
\def\d{{\rm d}}

%\input aa.cmm
%? This is the macro modification to suppress the bounding boxes of figures

%\refereelayout
\MAINTITLE{Equation of state and helioseismic inversions}
\AUTHOR{Sarbani Basu and J. Christensen-Dalsgaard}
\INSTITUTE{
Teoretisk Astrofysik Center, Danmarks Grundforskningsfond, and
Institute for Fysik og Astronomi, Aarhus Universitet,
DK-8000 Aarhus C, Denmark}
\OFFPRINTS{S. Basu}
\DATE{Received \ \ , accepted \ \ }

\ABSTRACT{
Inversions to determine the squared isothermal sound speed
and density  within the Sun often use the helium abundance $Y$ as the second
parameter.
This requires the explicit use of the equation of state (EOS),
thus potentially leading to systematic errors in the
results if the equations of state of the reference model 
and the Sun are not the same.
We demonstrate how this potential error can be suppressed.
We also show that it is possible to invert
for the intrinsic difference in the adiabatic exponent
$\Gamma_1$ between two equations of state.
When applied to solar data 
such inversion rules out the EFF equation of state completely,
while with existing data
it is difficult to distinguish between other equations of state.
}
\THESAURUS{09(06.15.1; 06.09.1; 02.05.2)}
\maketitle
\MAINTITLERUNNINGHEAD{ EOS and helioseismic Inversions}
\AUTHORRUNNINGHEAD{ S. Basu \& J. Christensen-Dalsgaard}

\titlea{Introduction}

Solar oscillation frequencies can be inverted  to deter\-mine the internal
structure of the Sun.
This is generally done by relating
the frequency differences between a solar model and the Sun 
to differences in the structure by linearising the oscillation
equations under the assumption of hydrostatic equilibrium. 
If, for example, we express the frequencies in terms of the
squared adiabatic sound speed $c^2$ and the density $\rho$,
the result can be written
$$
{\delta \omega_i \over \omega_i}
= \int  K_{c^2,\rho}^i{ \delta c^2 \over c^2}\d r +
\int K_{\rho,c^2}^i {\delta \rho\over \rho} \d r
 +{F_{\rm s}(\omega_i)\over E_i} \; .
\eqno\eqname\inveq
$$
({\eg} Dziembowski {\etal} 1990).
Here $\delta \omega_i$ is the difference
in the frequency $\omega_i$ of the $i$th mode between the
solar data and a reference model. 
The kernels $K_{c^2,\rho}^i$ and $K_{\rho,c^2}^i$
are known functions of the reference model which relate the changes in
frequency to the changes in $c^2$ and $\rho$ respectively.
The term in $\Fsurf$ results from the near-surface errors
in the model, such as the assumption of adiabatic oscillations;
$E_i$ is the inertia of the mode, normalised by the photospheric amplitude
of the displacement.

The kernels for the $(c^2,\rho)$ combination can be easily converted
to kernels for others pairs of variables like
$(\Gamma_1,\rho)$, $(u, \Gamma_1)$ with no extra assumptions (cf. Gough 1993);
here $u\equiv p/\rho$ is the squared isothermal sound speed,
$p$ being pressure.
However, very often in addition to the oscillation equations
it is assumed that the equation of state is known;
this may be used to construct the kernels
for $u$ or $\rho$ and $Y$. 
The implicit assumption that the equations of state in the Sun 
and reference model are the same leads  to potential systematic
errors in the inversion  for $u$ or $\rho$.

\titlea{Formulation of the inverse problem}

The conversion of the kernels for $(c^2, \rho)$ to those for $(u,Y)$ uses
${\delta \ln c^2}={\delta \ln \Gamma_1}+\delta \ln u$.
Had the equation of state been known, $\delta \Gamma_1$
could have been determined from
$\Gamma_1=\Gamma_1(p,\rho,\{X_i\})$, where $\{X_i\}$ is the composition.
In fact, $\delta \Gamma_1$ contains an
additional term, {\viz} the intrinsic difference 
$(\delta \Gamma_1/\Gamma_1)_{\rm int}$ at fixed $(p,\rho,\{X_i\})$
between the true and the model equations of state.
Characterizing the composition by the abundances by mass
$Y$ and $Z$ of helium and heavy elements
and assuming an unchanged heavy-element abundance $Z$, we therefore get
$$\eqalignno{
{\delta \Gamma_1\over \Gamma_1}=&
\left({\partial\ln \Gamma_1\over\partial Y}\right)_{p,\rho}\delta Y
+ {\left({\partial\ln \Gamma_1\over\partial\ln  p}\right )_{\rho, Y}}
{\delta p\over p} + \cr
& \left(\partial\ln \Gamma_1\over\partial\ln \rho\right )_{p,Y}
{\delta \rho\over \rho}
+ \left({\delta \Gamma_1\over \Gamma_1} \right )_{\rm int}
\;.&  \eqname\parti\cr}
$$
Note that differences in $Z$, or in the relative composition
of the heavy elements, will also appear in 
$(\delta \Gamma_1/\Gamma_1)_{\rm int}$.

Equation~{\parti} can now be used to rewrite Eq.~{\inveq}
in terms of $\delta u/u$ and $\delta Y$,
by expressing the terms in $\delta \rho/\rho$ and $\delta p/p$ in
terms of $\delta u/u$ by means of the equation of hydrostatic support.
This expression contains a con\-tri\-bu\-tion from the intrinsic
difference in $\Gamma_1$
weighted by the kernel for $c^2$ at constant $\rho$.
Thus, the full equation is:
$$
\eqalignno{
{\delta \omega_i \over \omega_i}
= & \int  K_{u,Y}^i{ \delta u \over u}\d r +
\int K_{u,Y}^i {\delta Y} \d r \cr + 
& \int  K_{c^2,\rho}^i
{ \left(\delta \Gamma_1 \over \Gamma_1\right)}_{\rm int}\d r
+{\Fsurf(\omega_i)\over E_i}\;. & \eqname\truly}
$$
In the expression normally used to invert for $(u, Y)$ the term in
${\left(\delta \Gamma_1 /\Gamma_1\right)}_{\rm int}$
is ignored; this clearly introduces a systematic error,
if the equations of state are in fact different.
A similar argument also holds for density in\-ver\-sions using the
pair $(\rho,Y)$. Here too the con\-tri\-bu\-tion from 
${\left(\delta \Gamma_1 /\Gamma_1\right)}_{\rm int}$ should
be taken into account.
We also note that using Eq.~\truly, one may directly invert for
${\left(\delta \Gamma_1 /\Gamma_1\right)}_{\rm int}$.

\titlea{Inversion technique}

We have used the Subtractive Optimally Localised
Avera\-\-ges method of Pijpers \& Thompson (1992),
adapted to inversion for structure differences
({\eg} Basu {\etal} 1996).
The principle of the inversion technique is to form linear
combinations of Eqs~\truly\ with weights $c_i(r_0)$ chosen such as to
obtain an average of, for example,
$\delta u/u$ localised near $r = r_0$
while suppressing the contributions from the remaining
terms in Eqs~{\truly}, including the near-surface errors.
In addition, the statistical errors in the combination must be constrained.

To invert for $\delta u/u$ the coefficients $c_i$ are chosen to minimise
$$\eqalignno{
&\int\left(\sum_i c_i K_{u,Y}^i -{\cal T}\right)^2\d r+ 
\beta_1\int \left(\sum_i c_i w(r)K_{Y,u}^i\right)^2\d r \cr
&+\beta_2\int \left(\sum_i c_i w(r)K_{c^2,\rho}^i\right)^2\d r
+\mu\sum_{i,j}c_ic_j E_{ij} \; ,& \eqname\funcc \cr}
$$
with the constraint that the averaging kernel be uni\-modu\-lar, i.e.,
$$
\sum_ic_i(r_0)\int_0^R K_{u,Y}^i(r)\d r =1 \; .
\eqno\eqname\unimod
$$
Here, ${\cal T}(r_0, r)$ is a target averaging kernel, chosen to be
a Gaussian of unit area centred at $r_0$.
$E_{ij}$ is the covariance matrix of errors in the data.
The parameters $\beta_1$ and $\beta_2$ control the contributions of 
$\delta Y$ and $(\delta\Gamma_1/\Gamma_1)_{\rm int}$, respectively,
and $\mu$ is a trade-off parameter
which controls the effect of data noise. The function $w(r)$ is a
suitably chosen, increasing function of radius, which ensures that the
contributions from the second  and third terms from the surface layers are
suppressed properly.

To reduce the influ\-ence of near-surface uncertainties we apply the
additional constraints that
$$
\sum_i c_i(r_0) E_i^{-1}\Phi_\lambda(\omega_i)=0 \; ,
\quad \lambda=0,\ldots,\Lambda \; ,
\eqno\eqname\sur
$$
where the $\Phi_\lambda$ are B-Splines
with a suitably scaled argu\-ment ({\cf} D\"appen {\etal} 1991).

To carry out an inversion for
$(\delta \Gamma_1/\Gamma_1)_{\rm int}$ one minimises instead
$$\eqalignno{
&\int\left(\sum_i c_i K_{c^2,\rho}^i -{\cal T}\right)^2\d r+ 
\beta_1\int \left(\sum_i c_i w(r)K_{u,Y}^i\right)^2\d r \cr
&+\beta_2\int \left(\sum_i c_i w(r)K_{Y,u}^i\right)^2\d r
+\mu\sum_{i,j}c_ic_j E_{ij} \; .& \eqname\funce \cr}
$$

We have used four solar models for this work. All models have
been constructed with OPAL opacities (Ig\-le\-sias, Rogers \& Wilson 1992)
at temperatures higher than $10^4$ K
and Kurucz tables (Kurucz 1991) at lower temperatures.
The models have been constructed with different equations of state ---
 Livermore (OPAL)
(Rogers, Swenson \& Iglesias, 1996)
 MHD  (e.g. Mihalas, D\"appen \& Hummer 1988),
 EFF (Eggleton, Faulkner \& Flannery 1993) and CEFF
(cf. \jcd\ \& D\" appen 1992).
The properties of the models, identified by the EOS,
are summarised in Table~1. Model OPAL is Model S of
Christensen-Dalsgaard et al. (1996).

\begtabfull
\tabcap{1}{Solar models used;
$d_{\rm CZ}$ is the depth of the convection zone,
and $T_{\rm c}$ and $\rho_{\rm c}$ are central temperature and density}
\table{\hsize}{#\hfil&&\hfil #\hfil\cr
\tabmidrule
\noalign{\medskip}
EOS&$Z/X$& $d_{\rm CZ}/R$& $T_{\rm c}$&$\rho_{\rm c}$\cr
  & & & $(10^6\hbox{ K})$ & $(\hbox{g/cm}^{3})$\cr
\noalign{\medskip}
\tabmidrule
\noalign{\medskip}
OPAL& 0.0245 & 0.2885 & 15.67& 154.2\cr
MHD & 0.0245 & 0.2876 & 15.67& 154.5\cr
CEFF& 0.0248 & 0.2863 & 15.68& 155.0\cr
 EFF& 0.0248 & 0.2852 & 15.74& 157.2\cr
\noalign{\medskip}
\tabmidrule}
\endtab
We use solar oscillation data obtained by the LOWL instrument during
the first year of data collection (Tom\-czyk et al. 1995,
Schou and Tom\-czyk, in preparation).
The dataset consists of modes
of degrees 0 to 99 in the frequency range 1 to 3.5 mHz.
The observed modeset and errors were also used in tests for solar models, 
in order to get realistic properties of the inversion.

\titlea{Results}

\begfig 4.8cm
\moveleft 0 true cm\vbox to 0 cm{\vskip -5.3 true cm
\epsfxsize=8.2 true cm\epsfbox{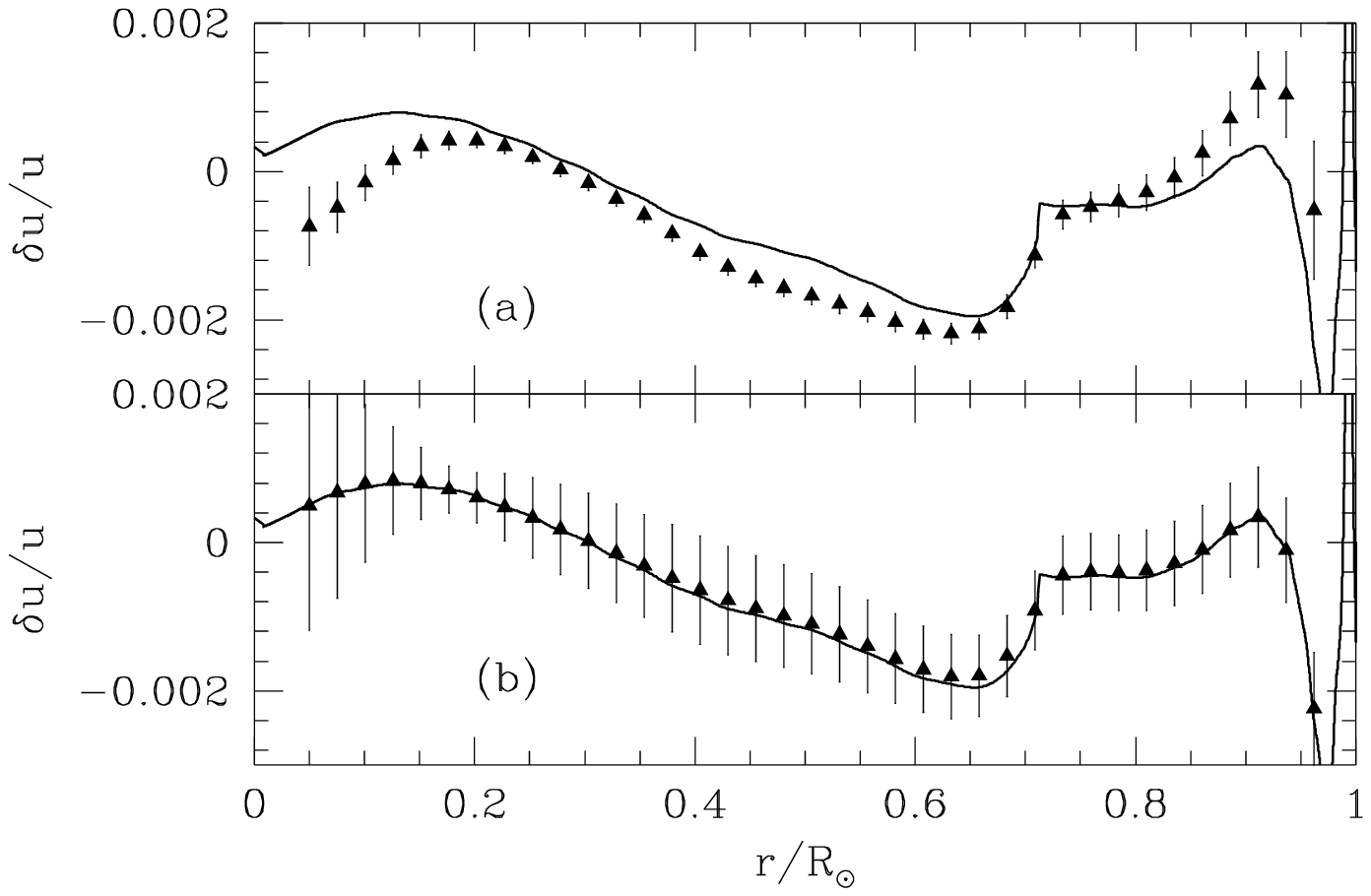}\vskip 0.3 true cm}
\figure{1}{The inversion for the squared isothermal sound speed ($u$)
difference between model MHD and model OPAL. 
The solid line is the exact difference and the points are the
difference obtained by inverting the frequency differences between
the models.
{\bf a} Inversion
results when the intrinsic difference in $\Gamma_1$ between the OPAL
and MHD equations of state is ignored. 
{\bf b } Inversion
results when the intrinsic difference is taken into account. The
vertical error-bars are 1$\sigma$ propagated errors }
\endfig

\begfig 4.8cm
\moveleft 0 true cm\vbox to 0 cm{\vskip -5.3 true cm
\epsfxsize=8.2 true cm\epsfbox{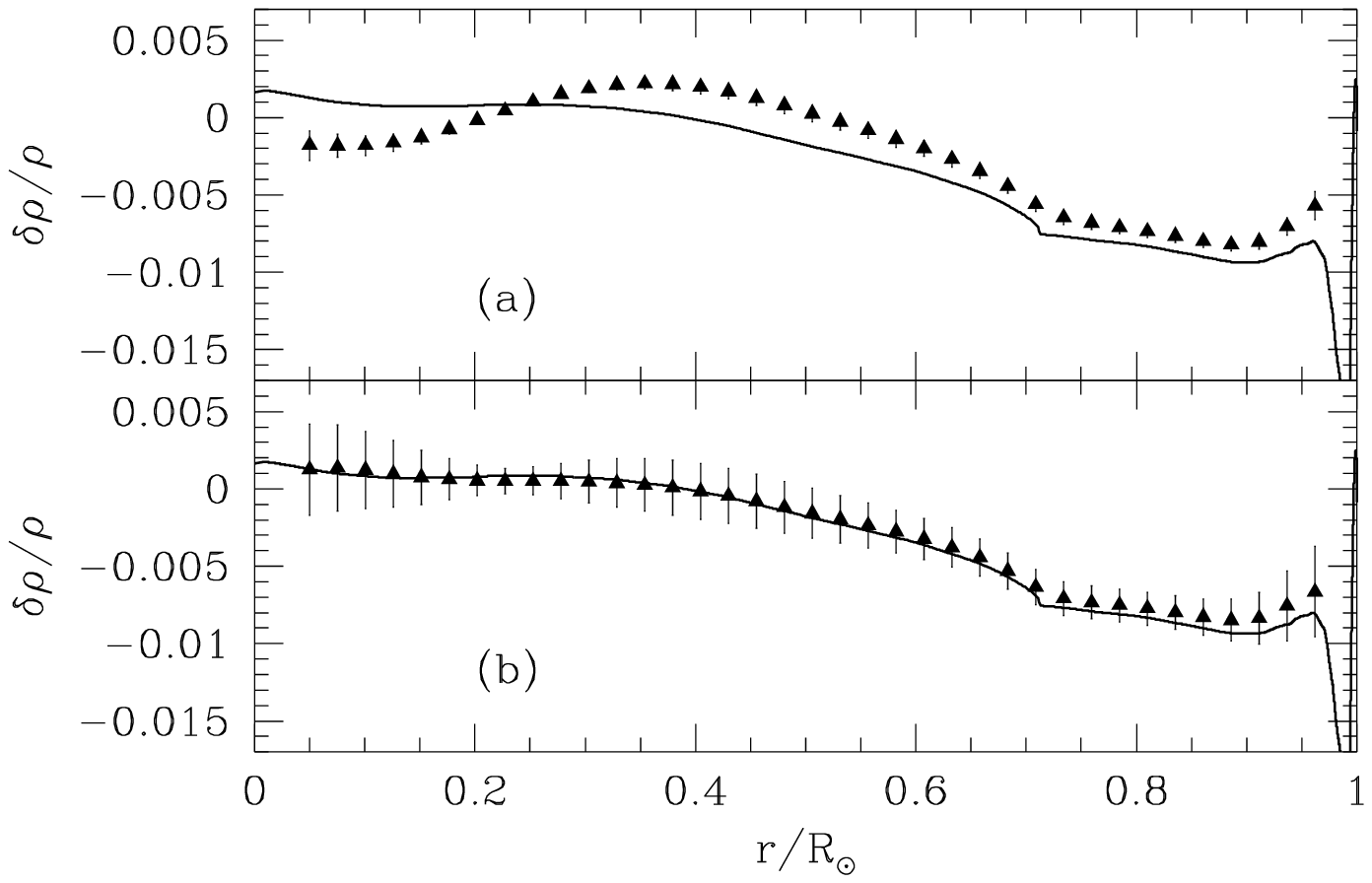}\vskip .5 true cm}
\figure{2}{The same as Fig.~1, but for the density difference
between models MHD and OPAL
}
\endfig

\begfig 6.2cm
\moveleft -.7 true cm\vbox to 0 cm{\vskip -6.6 true cm
\epsfxsize=7.8 true cm\epsfbox{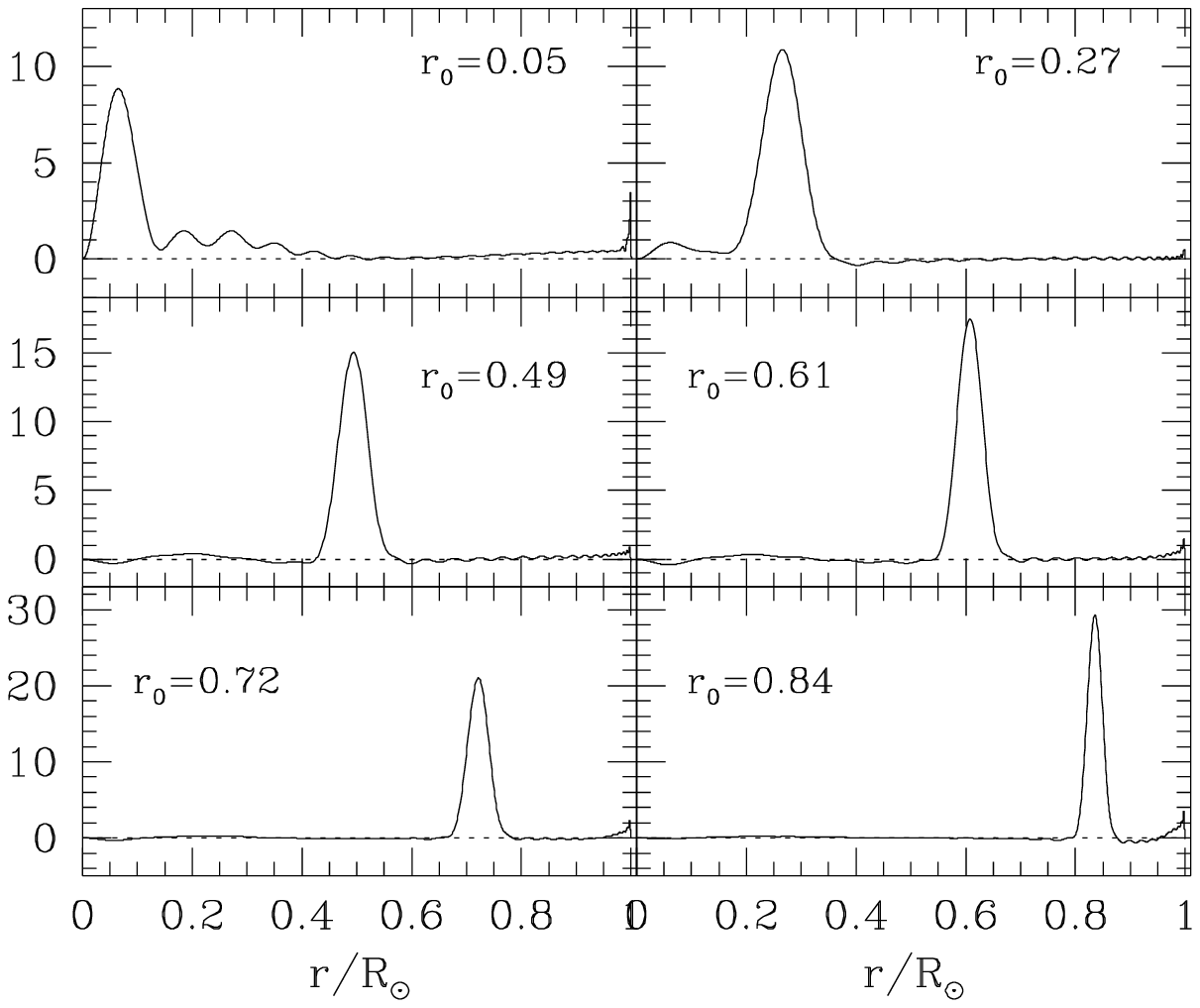}\vskip -0.0 true cm}
\figure{3}{Averaging kernels for the inver\-sion for the intrinsic
$(\delta\Gamma_1/\Gamma_1)_{\rm int}$
with the LOWL mode set and the OPAL model as the reference model
}
\endfig

\noindent
In Fig.~1 we show inversion results,
without and with suppression of 
${\left(\delta \Gamma_1 / \Gamma_1\right)}_{\rm int}$,
for $u$ in Model MHD with Model OPAL as the reference model. 
The resolution of the inversion is the same in both cases.
Note that
when the contribution from 
${\left(\delta \Gamma_1 / \Gamma_1\right)}_{\rm int}$ is not constrained,
the results are not very accurate.
The results improve dramatically, particularly in the core,
when the intrinsic difference is taken into account. 
However, the price paid for increased accuracy is decreased precision,
as reflected in the increased error-bars.
Indeed, it is evident that the use of the data to suppress
the possible error in the EOS  reduces the amount of information available
for the de\-ter\-mi\-na\-tion of $\delta u/u$ and hence results in
larger errors if the resolution is kept approximately the same.
The results for density inversion are shown in Fig.~2.
We note that the errors in the inversion corrected for
a possible inconsistency in the equation of state are much
reduced if accurate data on higher-degree modes are available,
as is the case, {\eg}, for the frequencies obtained by the
SOI/MDI experiment on SOHO (cf. Kosovichev et al. 1997).

Although it is useful to be able to suppress the effects
of errors in the equation of state when inverting for $u$,
it is evidently of greater interest to obtain a localized
measure of these errors, {\ie}, 
to invert for the intrinsic $(\delta \Gamma_1/\Gamma_1)_{\rm int}$ between
the equations of state of the Sun and the model.
To illustrate our ability to achieve such localization,
Fig.~3 shows averaging kernels for the in\-ver\-sions
for $(\delta \Gamma_1/\Gamma_1)_{\rm int}$ 
through minimization of expression~{\funce}.
Note that for 
$r_0 \gwig 0.5 R_\odot$ the averaging kernels are quite well
localized, indicating that reliable inversion is in fact possible.
The averaging kernels are not as small near the surface as
one would hope for;
inclusion of higher-degree modes  would substantially improve the
behaviour in this region.

\begfig 7.5cm
\moveleft 0 true cm\vbox to 0 cm{\vskip -8 true cm
\epsfxsize=8 true cm\epsfbox{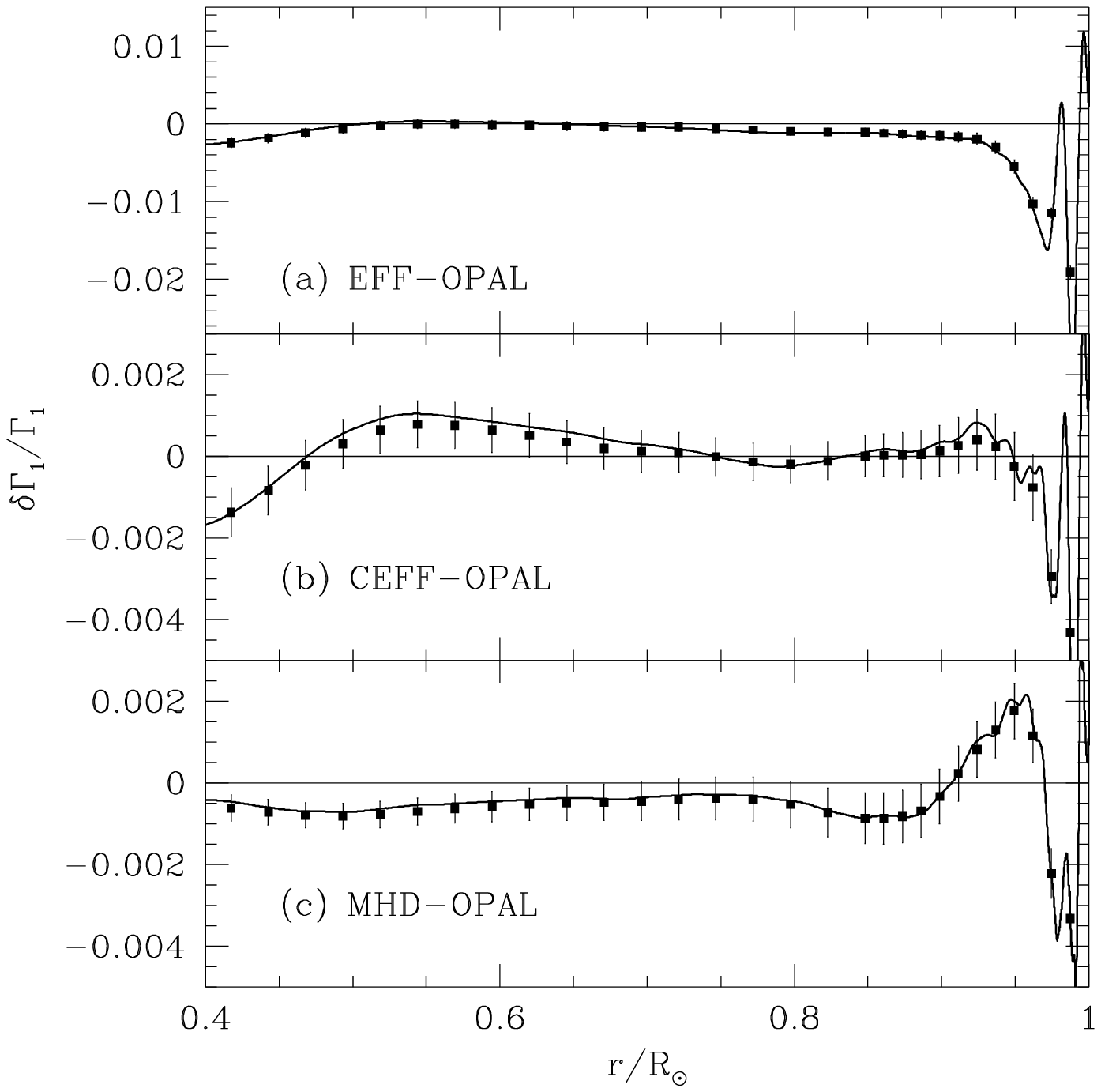}\vskip 0.5 true cm}
\figure{4}{The results of inversion for  the intrinsic $\Gamma_1$
difference between {\bf a} EFF, {\bf b} CEFF and {\bf c} MHD models and Model OPAL.
The solid line is the exact difference and the points are the
difference obtained by inverting the frequency differences between
the models.
Note that the scale in Panel {\bf a} is much larger than that
in Panels {\bf b} and {\bf c}}
\endfig
Fig.~4 shows the inversion for the intrinsic differences in $\Gamma_1$,
using the Models MHD, CEFF and EFF as test models
and Model OPAL as reference.
For comparison are shown exact differences resulting
from differences in the equation of state,
evaluated at fixed $p$, $\rho$, and $Y$ in Model OPAL.
The inversion of the frequencies clearly successfully reproduces
even the subtle intrinsic EOS differences
between the MHD and OPAL formulations,
although the statistical errors are fairly substantial
com\-par\-ed with these differences, at least beneath the dominant
ionization zones.

\begfig 7.5cm
\moveleft 0 true cm\vbox to 0 cm{\vskip -8 true cm
\epsfxsize=8.0 true cm\epsfbox{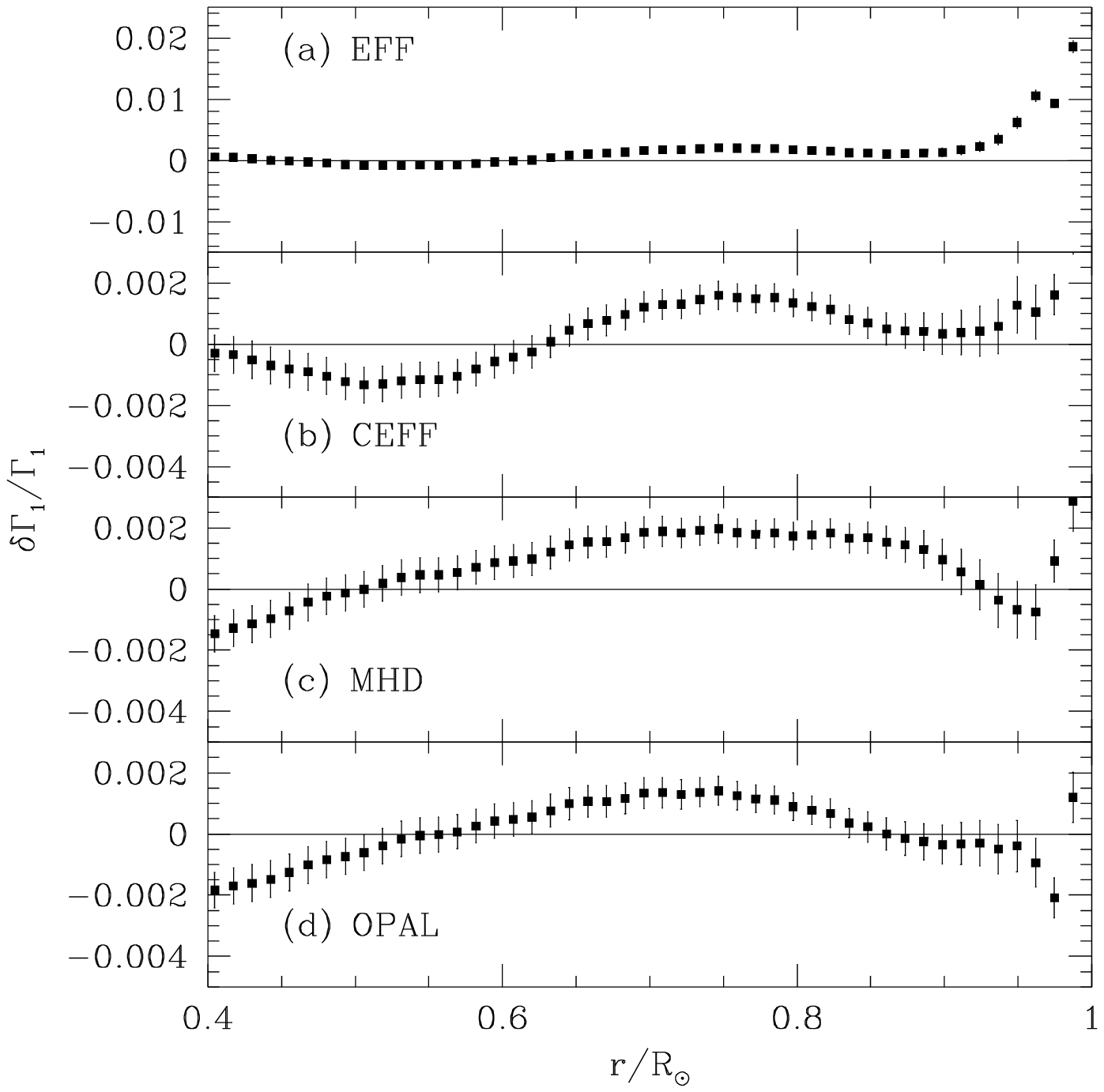}\vskip -0.0 true cm}
\figure{5}{The intrinsic $\Gamma_1$ difference between the Sun and the
EFF, CEFF, MHD and OPAL models obtained by inversion of LOWL
Year-1 data. Note the difference in scale between panel {\bf a} and the
other panels}
\endfig

Given the success of this test on artificial data,
we may consider differences between the solar and the model
equations of state, as obtained from analysis of the observed frequencies.
Fig.~5 shows the resulting intrinsic differences in $\Gamma_1$ between
the Sun and the four models of Table 1.
It is evident that the EFF equation of state
is inconsistent with the data.
With the current 
level of errors, it is difficult to distinguish between the other three
equations of state. 
Our ability to do so would be greatly improved by analysis
of higher-degree data,
since we may expect that the dominant differences in
the equations of state are close to the surface of the Sun.

\titlea{Conclusions}

We have shown that inversions for the squared 
isothermal sound speed $u$ and the density $\rho$ may suffer
from sys\-te\-ma\-tic errors when based on the common
implicit assumption that the equations of state
in the Sun and the reference model are the same.
These errors  can be 
removed by suppressing the contribution from the intrinsic difference
in $\Gamma_1$ to the frequency difference. 
However, this is achieved at the price of an increase in the propagated errors.

We also show that we can successfully invert for the intrinsic difference
in $\Gamma_1$ between the currently available equations of state.
This differs from the analysis by Elliott (1996) who
investigated the EOS in terms of the total difference
between the solar and the model $\Gamma_1$.
Inversions of solar oscillation fre\-quen\-cies show that the
EFF equation of state can be ruled out by direct inversions.
With the current level of data errors, it is difficult to judge the significance
of  the differences between the solar equation of state and the CEFF, MHD
and OPAL equations of state. 
We hope, however, that as more precise data and
data on high-degree modes become available, this method can be used as
a direct test of the solar equation of state.

\acknow{We thank M. J. Thompson for useful comments.
This work was supported  by the Danish Na\-tional Research
Foundation through its estab\-lish\-ment of the Theoretical Astrophysics
Center}

\begref{References}

\ref
Basu S., Christensen-Dalsgaard J., P\'erez Hern\'andez F., 
Thompson M.J., 1996, MNRAS 280, 651

\ref
Christensen-Dalsgaard J., \& D\" appen W., 1992, A\&AR 4, 267

\ref
Christensen-Dalsgaard J., D\"appen W., Ajukov S.V., et al., 1996,
Science 272, 1286

\ref
Dziembowski~W.A., Pamyatnykh A.A.,  Sienkiewicz R., 1990, MNRAS 244, 542

\ref
Eggleton P.P., Faulkner J., Flannery B.P., 1973, A\&A 23, 325

\ref
Elliott J.R., 1996, MNRAS 280, 1244 

\ref
Gough D.O., 1993, in:  Zahn J.-P., Zinn-Justin J., eds,
Astrophysical Fluid Dynamics: Les Houches, Session XLVII,
Elsevier, Amsterdam, p. 399

\ref
Iglesias C.A., Rogers F.J.,  Wilson B.G., 1992, ApJ 397, 717

\ref
Kosovichev A.G., Schou J., Scherrer P.H., et al., 1997, Sol. Phys., in 
press

\ref
Kurucz R. L., 1991,
in: Stellar atmospheres: beyond classical models,
Crivellari L., Hubeny I., Hummer D. G., eds,
NATO ASI Series, Kluwer, Dordrecht, p. 441

\ref
Pijpers  F. P., \& Thompson M. J., 1992, A\&A 262, L33

\ref
Mihalas D., D\"appen W. \& Hummer D. G., 1988, ApJ 331, 815

\ref
Rogers F.J., Swenson F.J., \&  Iglesias C.A., 1996, ApJ 456, 902

\ref
Tomczyk S., Streander K., Card G., Elmore D., Hull H., Caccani A.,
1995, Solar Phys.  159, 1 
\endref

\bye